\documentclass[10pt,journal,compsoc]{IEEEtran}
 
\usepackage{amsmath}
\usepackage{ntheorem}
\usepackage{algorithm}  
\usepackage{algorithmic}
\usepackage{graphicx}
\usepackage{multirow}
\usepackage{amssymb}
\usepackage{color,xcolor}
\usepackage{ulem}
\usepackage{bbm}
\ifCLASSOPTIONcompsoc
  \usepackage[nocompress]{cite}
\else
  \usepackage{cite}
\fi
\ifCLASSOPTIONcompsoc
 \usepackage[caption=false,font=footnotesize,labelfont=sf,textfont=sf]{subfig}
\else
 \usepackage[caption=false,font=footnotesize]{subfig}
\fi

\begin{document}
\title{Discovering Categorical Main and Interaction Effects Based on Association Rule Mining}
\author{Qiuqiang~Lin and Chuanhou~Gao,~\IEEEmembership{Senior Member,~IEEE}
\IEEEcompsocitemizethanks{\IEEEcompsocthanksitem Q. Lin and C. Gao are with the School of Mathematical Sciences, Zhejiang University, Hangzhou 310027, China.\protect\\
E-mail: \{11735034, gaochou\}@zju.edu.cn}
\thanks{Manuscript received June 5, 2020}}

\IEEEtitleabstractindextext{
\begin{abstract}
    With the growing size of data sets, feature selection becomes increasingly important. 
    Taking interactions of original features into consideration will lead to extremely high dimension, 
    especially when the features are categorical and one-hot encoding is applied.
    This makes it more worthwhile mining useful features as well as their interactions.
    Association rule mining aims to extract interesting correlations between items, 
    but it is difficult to use rules as a qualified classifier themselves.
    Drawing inspiration from association rule mining, we come up with a method that uses association rules to select features and their interactions, then modify the algorithm for several practical concerns.
    We analyze the computational complexity of the proposed algorithm to show its efficiency.
    And the results of a series of experiments verify the effectiveness of the algorithm.
\end{abstract}

}

\maketitle

\IEEEraisesectionheading{\section{Introduction}\label{sec:introduction}}
\IEEEPARstart{D}{ealing} with massive data is a common situation in contemporary scientific research and practical applications.
Whether one can extract useful information from data is critical to model's effectiveness and computational efficiency.
As both the number of features (denoted by $p$) and the number of samples (denoted by $n$) can be huge, it is usually computationally expensive, or intractable to fit a good model on original data without feature selection.

What's more, it is often the case that not only original features but also the interactions of input features count.
Even only the interactions of two features are considered, there will be $p(p+1)/2$ features in total.
As the number of interactions grows quadratically with the number of input features, it can be comparable with or larger than the sample size $n$, which makes the problem more challenging.

It is a well-established practice among statisticians fitting models on interactions as well as original features. 
There exist a number of works dealing with interactive feature selection, especially for pairwise interactions. 
Many of them select interactions by hierarchy principle \cite{bien13, hao14, hao18, agrawal19}. 
However, the theoretical analysis of these methods are based on the hierarchy assumption, 
so the practical performance may be unsatisfactory for the case where the assumption does not hold. 

There are also some works free of the hierarchy assumption. 
Thanei et al. propose the xyz algorithm, 
where the underlying idea is to transform interaction search into a closest pair problem which can be solved efficiently in subquadratic time \cite{Thanei18}.
And instead of the hierarchy principle, Yu et al. come up with the reluctant principle, which says that one should prefer main effects over interactions given similar prediction performance \cite{yu19}.

Most of the above-mentioned works concentrate on regression task and numerical features.
In fact, categorical features need interaction selection more eagerly than numerical features, 
since each categorical feature can have a huge number of different categories, 
which will naturally result in extremely high dimension if one-hot encoding is applied. 
Thus it's impossible to take all the pairwise interactions into consideration, let alone interactions of higher order.
Moreover, interactions of discrete features are usually more interpretable than continuous ones. 
For example, if $x$ means ``the temperature of yesterday'' and $y$ means ``the temperature of today'', it's hard to say what's the practical meaning of $xy$. On the contrary, if $x$, $y$ are two binary features and $x$=1 represents ``yesterday is sunny'', $y$=1 represents ``today is sunny'', then $xy$=1 means ``both yesterday and today are sunny'' and $xy$=0 means ``either yesterday or today is not sunny''. 
From the discussion above, it is worthwhile finding a method that can select useful interactions of categorical features.
To deal with this problem, Shah et al. come up with random intersection trees\cite{shah14} and Zhou et al. propose BOLT-SSI \cite{zhou19}.

Mining association rules between items from a large database is one of the most important and well researched topics of data mining. 
Association rule mining problem was first stated by Agrawal \cite{agrawal93, agrawal98}, and further studied by many researchers\cite{brin97, Savasere95, Han00, Zaki99, Hipp00}; 
see \cite{luna19} for more information about data mining.
Let $I=\{ i_1, i_2, ..., i_m\}$ be a set of items, and $X$, $Y$ are two subsets of $I$, called itemsets.
Association rule mining aims to extract rules in the form of ``$X\to Y$'', where $X\cap Y=\emptyset$. 
Calling $X$ the antecedent and $Y$ the consequent, the rule means $X$ implies $Y$.
The support of an itemset $X$ is the number of records that contain $X$.
For an association rule ``$X\to Y$'', its support is defined as the fraction of records that contain $X\cup Y$ to the total number of records in the database,  
and its confidence is the number of cases in which the rule is correct relative to the number of cases in which it is applicable, or equivalently, support($X\cup Y$)/support($X$).

Originally association rule mining is designed for extracting interesting correlations, frequent patterns, associations or casual structures among sets of items in the transaction databases or other data repositories.
For classification tasks with discrete input features, we can regard ``the input feature has a specific value'' and ``the label has a specific value'' as items, and use association rule mining methods to extract several association rules which have ``the label has a specific value'' in consequence. The rules in this form are called ``class association rules''\cite{liu98}. 
Based on this idea, there are lots of works that mine a set of association rules and use them as a classifier\cite{liu98, dong99, Meretakis99, liu00, li01, yin03}, and the classification approach in this form is called ``associative classification''. 
The problem is that these methods need to mine a huge number of rules to cover different antecedents directly, which is in lack of efficiency and hard to handle complicated relationships.

Alternatively, we could use association rules to help selecting useful features for another classification model rather than classify instances themselves. For a class association rule, its antecedent is likely corresponding to a meaningful feature for the prediction task. 
If the antecedent contains exactly one element, it corresponds to an original feature (called main effect in some literature), otherwise it corresponds to an interactive feature. 
This idea sheds some light on interactive feature selection. 
We could first mine the association rules, then transform the antecedents to features. 
For instance, if an association rule is ``$(X_i=x_i)\to (Y=c)$'', then we can generate a binary feature named ``$X_i=x_i$'' that will be assigned as 1 if and only if $X_i=x_i$, which seems like one-hot encoding. 
Similarly, if an association rule is ``$(X_i=x_i, X_j=x_j)\to (Y=1)$'', then we can generate a binary feature named ``$X_i=x_i, X_j=x_j$'', which will be assigned as 1 if and only if $X_i=x_i$ and $X_j=x_j$.
The detail of how to generate features from a set of association rules is given in Algorithm \ref{ar2f}.

Our main contributions in this paper are listed below:
\begin{itemize}
    \item Propose a principle of feature selection, which we call the information principle.
    \item Come up with an algorithm for feature selection and interaction inspired by association rule mining methods, and modify it for several practical concerns. 
    \item Analyze the time and space complexity of the proposed algorithms, based on which we give some suggestions on parameter selection.
    \item Conduct a series of experiments to verify the effectiveness of the proposed algorithms.
\end{itemize}

The rest of the paper is organized as follows. In Section~\ref{sec:principle}, we introduce three principles used in our algorithms. The first two principles are proposed in earlier related works, namely the hierarchy principle and the reluctant principle. Then we propose a principle of feature selection, which we call the information principle. In Section~\ref{sec:algorithm}, we formally introduce the algorithm for interactive feature selection by integrating association rule mining methods. A theoretical analysis of its space and time complexities is then given in Section~\ref{sec:complexity}, where we also give some suggestions on parameter selection. 
And some extensions of the proposed algorithm are given in Section~\ref{sec:modification}.
In Section~\ref{sec:experiment} we report the results of a series of experiments to verify the effectiveness of the algorithms.
Finally Section~\ref{sec:conclusion} concludes this paper.

\section{Principles}\label{sec:principle}
\subsection{Hierarchy Principle}
A main difficulty of making use of interactive features is that one should fit a model on $O(p^2)$ features, while only a fraction of the features really count. Taking all the interactions into consideration not only leads to heavy computational burden, but also makes the model hard to understand.
Hierarchy assumption is widely used in statistical analysis for interaction selection to avoid this difficulty\cite{bien13, hao14, hao18, agrawal19}.

\emph{\textbf{Hierarchy assumption}: An interaction effect is in the model only if either (or both) of the main effects corresponding to the interaction are in the model.}

Based on this assumption, an interaction is allowed to be added into the model only if the corresponding main effects are also in the model, which can tremendously reduce the computation cost.
For example, 
Bien et al. add a set of convex constraints to the lasso that honor the hierarchy restriction\cite{bien13};
Hao et al. tackle the difficulty by forward-selection-based procedures\cite{hao14};
Hao et al. consider two-stage LASSO and a new regularization method named RAMP to compute a hierarchy-preserving regularization solution path efficiently\cite{hao18}.
Agrawal et al. propose to speed up inference in Bayesian linear regression with pairwise interactions by using a Gaussian process and a kernel interaction trick \cite{agrawal19}.
Hierarchy is a well-established practice among statisticians, but disappointingly, the practical performance and theoretical analysis are only guaranteed when the hierarchy assumption holds, which is not always the case.

\subsection{Reluctant Principle}
Yu et al. get rid of the hierarchy assumption, and come up with the reluctant principle instead\cite{yu19}.

\emph{\textbf{The reluctant interaction selection principle}: One should prefer a main effect over an interaction if all else is equal.}

According to Yu et al., there are at least two reasons for preferring main effects. 
The first is that main effects are easier to interpret than interactions. 
Thus when presented with two models that predict the response equivalently, we should favor the one that relies on fewer interactions. 
The other reason is prioritizing main effects can lead to great computational savings both in terms of time and memory, since the total number of main effects is far smaller than the number of interactions.
Both the hierarchy principle and the reluctant principle can simplify the search of interactions, while the reluctant principle does not explicitly tie an interaction to its corresponding main effects.


\subsection{Information Principle}
For a discrete random variable, a criterion is needed to decide whether it is valuable for the task.
In other words, we have to evaluate the information we can receive by observing a specific value for this variable.
The amount of information of knowing the value of a variable can be viewed as the ``degree of surprise''\cite{bishop07}. 
If we are told that a highly improbable event has just occurred, we will have received more information than if we were told that some very likely event has just occurred, and if we knew that the event was certain to happen we would receive no information. 
Thus the larger difference between the posterior distribution and the prior distribution (of the response), the more information this observation contains. 

\emph{\textbf{Information principle}: Select the features if observing them makes largest difference between the posterior distribution and the prior distribution of the response.}

According to the information principle, it is the features which lead to the largest difference between posterior probabilities and prior probabilities that are meaningful, rather than the features corresponding to the largest(or smallest) posterior probabilities.
For categorical feature selection, dropping the features whose corresponding posterior probabilities of response is closest to the prior probabilities can preserve the information as much as possible.
Or from the other perspective, we can select the features whose corresponding distance of the posterior probability and the prior probability is largest.

\section{Algorithms}\label{sec:algorithm}
\subsection{Apriori Algorithm}
Apriori algorithm was first proposed by Agrawal\cite{agrawal98}, which is based on the concept of a prefix tree. 
Discovering association rules can be decomposed into two subproblems: first find the frequent itemsets, which have support above minimum support(denoted by minsupp), and then use these frequent itemsets to generate confident rules, which have confidence above minimum confidence(denoted by minconf).

The algorithm exploits the observation that a superset of an infrequent set can not be frequent. Thus when discovering frequent itemsets, only the itemsets whose every subset is frequent need to be examined. The detail of Apriori algorithm is given in Algorithm \ref{apriori}.

The main idea of Apriori algorithm is that only the itemset containing no infrequent itemsets need examining whether it is frequent. Therefore the number of candidate itemsets is reduced rapidly. This is very similar with the hierarchy principle, where only the interactions of selected main effects are possible to be chosen. However, the hierarchy assumption is kind of unnatural since it does not always hold. On the contrary, the idea of Apriori algorithm is guaranteed theoretically.

\begin{algorithm}
    \caption{GenerateFeatures}
    \label{ar2f}
    \begin{algorithmic}[1]
    \REQUIRE AR(a set of association rules);
    \ENSURE $F$(a set of features);
    \STATE{$F=\emptyset$}
    \FORALL{R in AR}
        \IF{R has the form of ``$(X_i=x_i)\to (Y=c)$''}
        \STATE {$F\leftarrow F\cup\{(X_i=x_i)\}$}
        \ELSIF{R has the form of ``$(X_i=x_i, X_j=x_j)\to (Y=c)$''}
        \STATE{$F\leftarrow F\cup\{(X_i=x_i, X_j=x_j)\}$}
        \ENDIF
    \ENDFOR
    \STATE{Return $F$}
    \end{algorithmic}
\end{algorithm}

\begin{algorithm}
    \caption{Apriori Algorithm}
    \label{apriori}
    \begin{algorithmic}[1]
    \REQUIRE D(database); \\
             minsupp(minimum support); \\
             minconf(minimum confidence);
    \ENSURE $R_t$(all association rules)
    \STATE {$L_1$= \{frequent 1-itemsets\}} 
    \FOR{$k=2$; $L_{k-1}\neq \emptyset$; k++}
        \STATE $C_k$ =apriori-gen($L_{k-1}$);  
        \FORALL{transactions $T\in D$}
            \STATE{$C_t$=subset($C_k$,T);}  
            \FORALL{candidates $C \in C_t$}
                \STATE{Count(C)$\leftarrow$Count(C)+1;}  
            \ENDFOR
        \ENDFOR
    \STATE{$L_k=\{ C\in C_t|{\rm Count(C)}\ge {\rm minsupp}\cdot|D|\}$}
    \ENDFOR
    \STATE{$L_f=\bigcup_k{L_k}$}
    \STATE{Return $R_t$={\rm GenerateRules}($L_f$, minconf)}
    \end{algorithmic}
\end{algorithm}

\subsection{ARAF: Association Rules as Features}
In this section, we describe a new method for using association rules as features, ARAF for short.
The association rules in this section refer specifically to class association rules. 
If there is exactly one item in the antecedent, the corresponding feature is a main effect, otherwise it's an interaction. 
An advantage of ARAF is there are no limitations on the size of the antecedents in principle, but we only consider main effects and pairwise interactions currently for simplification. Such a consideration is a common situation in the work related to interactions selection \cite{bien13,hao14,hao18,agrawal19,Thanei18,yu19}. Another advantage of association rules is their interpretability. 
Rules may be one of the most interpretable models for human beings, and the features generated by rules succeed the interpretability.
What's more, since ARAF is model-agnostic, practitioners has the freedom of choosing their favorite model using main effects and interactions selected by ARAF methods.

The most direct way of using association rules as features is that run a standard Apriori algorithm first, then transform the antecedents of the found rules to features. This algorithm is shown in Algorithm \ref{araf_standard}.
\begin{algorithm}
    \caption{ARAF-sa: ARAF with standard Apriori}
    \label{araf_standard}
    \begin{algorithmic}[1]
    \REQUIRE D(database); \\
    minsupp(minimum support); \\
    minconf(minimum confidence);
    \ENSURE $F$(generated features);
    \STATE {Run Apriori algorithm to get $FS_1$ and $FS_2$, where\\
    $FS_1$=\{S$\mid$S=($X_i$=$x_i$, Y=c), supp(S)$\ge$ minsupp\},\\
    $FS_2$=\{S$\mid$S=($X_i$=$x_i$, $X_j$=$x_j$, Y=c), supp(S)$\ge$ minsupp\}.}
    \STATE{Generate the confident rules from $FS_1$ and $FS_2$.\\
    $CR_1$=\{R=``($X_i$=$x_i)\to$ Y=c''$\mid$conf(R)$\ge$minconf, \\($X_i$=$x_i$, Y=c)$\in FS_1$\},\\
    $CR_2$=\{R=``($X_i$=$x_i, X_j$=$x_j)\to$ Y=c''$\mid$conf(R)$\ge$minconf, \\($X_i$=$x_i$, $X_j$=$x_j$, Y=c)$\in FS_2$\}.}
    \STATE{Return $F={\rm GenerateFeatures}(CR_1\cup CR_2)$}
    \end{algorithmic}
\end{algorithm}

The advantage of this method is that one can choose the thresholds, namely minsupp and minconf, manually to balance generality and accuracy of the selected rules.
But this also causes some problems that it is difficult to choose appropriate parameters. If minsupp or minconf is too large, few or no association rules will be generated. On the other hand, if minsupp or minconf is too small, the generated rules can be unreliable. Another problem of small parameters is that too many rules may occur, which makes it hard to guarantee space and time efficiency.

To avoid the inefficiency of the standard Apriori, we use the number of frequent itemsets and confident rules rather than thresholds to decide whether an itemset is frequent or a rule is confident. 
This approach is standard in interaction modeling approaches \cite{Thanei18, yu19}.
If there are too many itemsets having the same support or rules having the same confidence to keep all of them, we give priority to those occur earlier. 
This seems somewhat alternative, but coincides with the reluctant principle since an itemset always occurs earlier than its supersets during the procedure of Apriori.
Compared to setting thresholds beforehand, this method is more robust. What's more, by using data structure such as min-heap, the space complexity and time complexity can be analyzed. The modified algorithm is shown in Algorithm \ref{araf_fs}.

\begin{algorithm}
    \caption{ARAF-fs: ARAF with fixed size}
    \label{araf_fs}
    \begin{algorithmic}[1]
    \REQUIRE D(database); \\
    $d_{\rm freq}$(the number of frequent itemsets);\\ 
    $d_{\rm conf}$(the number of confident rules);
    \ENSURE $F$(generated features);
    \STATE{Count the support of itemsets in $IS_1$, where\\
           $IS_1$=\{($X_i$=$x_i$, Y=c), for $i\in$\{1,2,...,p\}\}}
    \STATE{Generate $FS_1$=\{$d_{\rm freq}$ most frequent itemsets in $IS_1$\}}
    \STATE{Count the support of itemsets in $IS_2$, where\\
           $IS_2$=\{($X_i$=$x_i$, $X_j$=$x_j$, Y=c)$\mid$\\
           ($X_i$=$x_i$, Y=c), ($X_j$=$x_j$, Y=c)$\in$ $FS_1$\}}
    \STATE{Generate $FS_2$=\{$d_{\rm freq}$ most frequent itemsets in $IS_1 \cup IS_2$\}}
    \STATE{Calculate the confidence of rules in $AR$, where\\
           AR=\{class association rules generated from $FS_2$\}}
    \STATE{Generate $CR$=\{$d_{\rm conf}$ most confident rules in $AR$\}}
    \STATE{Return $F$=GenerateFeatures(CR)}
    \end{algorithmic}
\end{algorithm}

Algorithm \ref{araf_fs} is theoretically equivalent to Algorithm \ref{araf_standard}, thus it has the same advantage as the latter.
But it is easier to tune the parameters of Algorithm \ref{araf_fs}, and space complexity as well as time complexity can be controlled by the parameters.
However, there are still some problems. For example, practical data is usually not balanced. When selecting frequent itemsets from an unbalanced data set, it's likely to choose the itemsets whose label is the major class. 
Therefore most of the generated rules have major class as consequence while we usually concern more about the minor class. 
To avoid this, it's worth forcing the algorithm to attach more importance to the minor class. 
By separately selecting the frequent itemsets for different classes, it's hopeful to find itemsets corresponding to every class. 

Another observation is when the data is unbalanced, the rule for major class(that is, the class with large prior probability) will usually have larger confidence(in other words, posterior probability).
So if we choose rules according to confidence, the rules for minor classes are likely to be overlooked. 
For binary classification, this may be insignificant since after identifying the instances of major class, the rest can be simply treated as the minor class. 
However, if there are several minor classes, we can hardly distinguish them from each other.
According to the information principle, it's the distance between posterior distribution and prior distribution rather than posterior distribution itself that matters. 
Thus for unbalanced multiclass classification tasks, confidence is not an ideal criterion.
Lift, the ratio of the confidence and the support of the consequent, is an alternative choice.
But it usually overcompensates, since a slight improvement for a minor class can lead to a large Lift.
Therefore, we make another try and define relative confidence of an association rule as 

\begin{equation}
    \label{rconf}
    \begin{aligned}
        {\rm rconf}(X\!\to \!Y)&=\frac{{\rm conf}(X\to Y)}{1-{\rm conf}(X\to Y)}/\frac{P(Y)}{1-P(Y)}\\
        &=\frac{{\rm supp}(X\cup Y)}{{\rm supp}(X)-{\rm supp}(X\cup Y)}\cdot\frac{n-{\rm supp}(Y)}{{\rm supp}(Y)}
    \end{aligned}
\end{equation}
where $P(Y)$ is the proportion of $Y$ in the database. 
This idea is also touched by some earlier works, such as \cite{yan09}, where a different ``relative confidence'' was defined.
In consideration of numerical stability, we add a small value to the denominator of (\ref{rconf}) in actual calculation. 
On the basis of this criterion, we come up with Algorithm \ref{araf_ub}, selecting the rules with largest relative confidence.

\begin{algorithm}
    \caption{ARAF-ud: ARAF for unbalanced data}
    \label{araf_ub}
    \begin{algorithmic}[1]
    \REQUIRE D(database); \\
    $d_{\rm freq}$(the number of frequent itemsets);\\ 
    $d_{\rm conf}$(the number of frequent rules);
    \ENSURE $F$(generated features);
    \STATE Count the support of itemsets in $IS_1^{(c)}$, where\\
    $IS_1^{(c)}$=\{($X_i$=$x_i$, Y=c), for $i\in$\{1,2,...,m\}\}, for $c \in C$, \\
    where C is the set of labels.
    \STATE Generate the set of frequent 1-itemsets for each class:\\
    $FS_1^{(c)}$=\{$\frac{d_{\rm freq}}{|C|}$ most frequent itemsets in $IS_1^{(c)}$\}\\ for $c \in C$.
    \STATE Count the support of itemsets in $IS_2^{(c)}$, where\\
    $IS_2^{(c)}$=\{($X_i$=$x_i$, $X_j$=$x_j$, Y=c)$\mid$\\($X_i$=$x_i$, Y=c), ($X_j$=$x_j$, Y=c)$\in$ $FS_1^{(c)}$\}
    \STATE Generate  the set of frequent itemsets for each class:\\
    $FS_2^{(c)}$=\{$\frac{d_{\rm freq}}{|C|}$ most frequent itemsets in $IS_1^{(c)} \cup IS_2^{(c)}$\}\\ for $c \in C$.
    \STATE{Calculate the relative confidence of rules in $AR$, where\\
    AR=\{class association rules generated from $\bigcup_{c}{FS_2^{(c)}}$\}}
    \STATE{Generate the set of confident class association rules\\
    $CR$=\{$d_{\rm conf}$ most relatively confident rules in $AR$\}}\\
    \STATE{Return $F$=GenerateFeatures(CR)}
    \end{algorithmic}
\end{algorithm}

There is still one question not settled, that it tends to obtain many redundant rules when using Algorithm \ref{araf_ub} to select association rules.
For example, if $X_1$=1 always holds and  ``($X_2$=1)$\to$(Y=1)'' is a good rule that has high frequency and confidence. 
Then ``$(X_1$=1, $X_2$=1)$\to$(Y=1)'' will have the same frequency and confidence as ``($X_2$=1)$\to$(Y=1)'', even though ``$X_1$=1'' has nothing to do with ``$Y$=1''. 

To overcome this difficulty, we adopt the reluctant principle. When the main effect is already chosen, we are reluctant to choose an interaction unless it brings us more information. That is, only the interactions that are more (relatively) confident than corresponding main effects will be chosen. To do this, when selecting (relatively) confident rules with 2-item antecedent in step 5, only those have larger (relative) confidence than their main effects will be selected. Noticing that the record contains an interaction will also contain corresponding main effects, which means main effects always have larger support than their interactions. Thus sorting the frequent itemsets by their support can ensure main effects occur earlier than their interactions.

The procedure of generating association rules without redundancy is shown in Algorithm \ref{araf_reluctant}. By utilizing the reluctant principle, the unnecessary interactions will be kept out, thus we can obtain more useful rules without increasing the number of frequent itemsets and confident rules. 

\begin{algorithm}
    \caption{ARAF: using association rules as features}
    \label{araf_reluctant}
    \begin{algorithmic}[1]
        \REQUIRE D(database); \\
        $d_{\rm freq}$(the number of frequent itemsets);\\ 
        $d_{\rm conf}$(the number of confident rules);
        \ENSURE $F$(generated features);\\
        {All the other steps are the same as Algorithm \ref{araf_ub}, except generating AR(in Step 5 there) by following steps.}
        \STATE{For each c $\in$ C, sort $FS_2^{(c)}$ by support}
        \STATE{Initialize $AR=\emptyset$}
        \FORALL{c $\in$ C}
            \FORALL{$I$ in $FS_2^{(c)}$}
            \IF{$I\in IS_1^{(c)}$}
            \STATE{Suppose $I$=$(X_i=x_i,Y=c)$\\
                   Denote $R_i$=``$(X_i=x_i)\to (Y=c)$''}
            \STATE{$AR\leftarrow AR\cup \{R_i\}$}
            \ELSIF{$I\in IS_2^{(c)}$}
            \STATE{Suppose $I$=$(X_i=x_i,X_j=x_j,Y=c)$\\
                   Denote $R_i$=``$(X_i=x_i)\to (Y=c)$'', \\
                   $R_j$=``$(X_j=x_j)\to (Y=c)$'',\\
                   $R_{ij}$=``$(X_i=x_i, X_j=x_j)\to (Y=c)$''}
            \IF{$R_i\notin AR$ or rconf($R_{ij}$)$\ge$rconf($R_i$)}
            \IF{$R_j\notin AR$ or rconf($R_{ij}$)$\ge$rconf($R_j$)}
            \STATE{$AR\leftarrow AR\cup \{R_{ij}\}$}
            \ENDIF
            \ENDIF
            \ENDIF
            \ENDFOR
        \ENDFOR
        \STATE{Return AR}
    \end{algorithmic}
\end{algorithm}

\section{Complexity Analysis and Parameter Selection}\label{sec:complexity}
In this section, the space complexity and time complexity of the proposed algorithms are analyzed. First we analyze the complexities of Algorithm \ref{araf_fs}, and then show that the modifications in Section~\ref{sec:algorithm} won't carry extra computation burdens. After that, some advise on tuning parameters are provided.

\subsection{Computational Complexity}
\label{sec:computational_complexity}


Here are some notations, $p$ is the number of categorical features, $n$ is the size of the database, $|C|$ is the number of different categories in label, $d_{\rm freq}$ is the number of frequent itemsets, $d_{\rm conf}$ is the number of confident rules, $d_{\rm conf}\le d_{\rm freq}$. 

$p$, $n$ and $|C|$ are determined by the data, while $d_{\rm freq}$ and $d_{\rm conf}$ are tuning parameters.
Assume the number of different categories of each discrete feature as well as $|C|$ is known in advance, in other words, they are constants independent of sample size $n$.

Let us focus on Algorithm \ref{araf_fs} first.

In Step 1, a traversal through the database is needed and we should record how many times ($X_i$=$x_i$, Y=c) occurs in the database. According to the assumption, the possible value of $X_i$ and $Y$ are constants known in advance so there are $O(p)$ different tuples in this form. Thus the time complexity is $O(np)$ and space complexity is $O(p)$.

In Step 2, we should find $d_{\rm freq}$ tuples with largest support in Step 1, which is a standard Top-K problem. By taking advantage of a min-heap of size $d_{\rm freq}$, the time complexity is $O(p\log (d_{\rm freq}))$ and space complexity is $O(d_{\rm freq})$.

In Step 3, we have to count how many times ($X_i$=$x_i$, $X_j$=$x_j$,Y=c) occurs in the database. Again, a traversal through the database is needed, and we should check $O(d_{\rm freq}^2)$ combinations for every instance. Therefore the time complexity is $O(nd_{\rm freq}^2)$ and space complexity is $O(d_{\rm freq}^2)$.

In Step 4, we should find $d_{\rm freq}$ tuples with largest support from the sets generated in Step 1 and Step 3. To do this, We could simply push the tuples found in Step 3 into the min-heap already used in Step 2. The time complexity is $O(d_{\rm freq}^2\log (d_{\rm freq}))$ and no extra space is required.

In Step 5, we can make use of (\ref{cal_conf}) to calculate the confidence of a rule.

\begin{equation}
    \label{cal_conf}
    {\rm conf}((X\!=\!x)\to(Y\!=\!c))=\frac{{\rm supp}(X\!=\!x, Y\!=\!c)}{\sum_{c'}{{\rm supp}(X\!=\!x, Y\!=\!c')}}
\end{equation}
where ($X$=$x$) could have the form of ($X_i$=$x_i$) or ($X_i$=$x_i, X_j$=$x_j$).

So we don't need new traversals. If the supports are stored in an associative array, the time complexity of inquiring the supports and calculate the confidence for a rule is $O(1)$, so we need $O(d_{\rm freq})$ time and space in total.

In step 6, we should generate $d_{\rm conf}$ most confident association rules from the rule sets gotten in Step 5. This is a Top-K problem again, we can use a min-heap of size $d_{\rm conf}$ to select the most confident rules from $d_{\rm freq}$ rules. Thus the time complexity is $O(d_{\rm freq}\log (d_{\rm conf}))$ and space complexity is $O(d_{\rm conf})$. 

Finally in step 7, we generate a feature from an association rule by simply extracting its antecedent, so $O(d_{\rm conf})$ time and space is required.

\begin{table}[!t]
    \renewcommand{\arraystretch}{1.3}
    \caption{Computational complexities of each step}
    \label{complexities}
    \centering
    \begin{tabular}{c||c|c}
        \hline
        Step & Time Complexity & Space Complexity \\
        \hline
        1 & $O(np)$ & $O(p)$ \\
        2 & $O(p\log (d_{\rm freq}))$ & $O(d_{\rm freq})$ \\
        3 & $O(nd_{\rm freq}^2)$ & $O(d_{\rm freq}^2)$ \\
        4 & $O(d_{\rm freq}^2\log (d_{\rm freq}))$ & $O(d_{\rm freq})$ \\
        5 & $O(d_{\rm freq})$ & $O(d_{\rm freq})$ \\
        6 & $O(d_{\rm freq}\log (d_{\rm conf}))$ & $O(d_{\rm conf})$\\
        7 & $O(d_{\rm conf})$ & $O(d_{\rm conf})$\\
        \hline
    \end{tabular}
\end{table}

The computational complexities of each step are summarized in Table \ref{complexities}. Noticing that $d_{\rm conf}\le d_{\rm freq}$, the time complexity and space complexity are:
\begin{itemize}
    \item Time: $O(np+p\log (d_{\rm freq})+nd_{\rm freq}^2+d_{\rm freq}^2\log (d_{\rm freq}))$
    \item Space: $O(p+d_{\rm freq}^2)$
\end{itemize}

The difference between Algorithm \ref{araf_ub} and Algorithm \ref{araf_fs} is that Algorithm \ref{araf_ub} uses different min-heaps for different target classes when selecting frequent sets, and calculate relative confidence instead of confidence for an association rule. Only Steps 2, 4 and 5 have changed. In Step 2, the total number of the frequent sets is $O(|C|\cdot\frac{d_{\rm freq}}{|C|})=O(d_{\rm freq})$, so space complexity remains. And for each itemset, Algorithm \ref{araf_ub} first use another $O(1)$ time to find the corresponding min-heap then push the itemset in, so the time complexity also stays the same. The analysis of Step 4 is similar. In Step 5, calculating relative confidence need only substitute the corresponding confidence and frequency into (\ref{rconf}), which brings no heavier computation. 

If we generate the set of association rules following the process of Algorithm \ref{araf_reluctant}, there are two additional operations. The first is sorting the frequent itemsets by their support before pushing them into the min-heap, and the next is checking whether a corresponding main effect with higher (relative) confidence is already in the min-heap when meeting an interaction. Applying the quick sort method\cite{Cormen09}, the time complexity of sorting is $O(d_{\rm freq}\log (d_{\rm freq}))$. There are $d_{\rm freq}$ frequent itemsets, and each corresponds to a class association rule. For every association rule, looking over the min-heap of confident rules needs $O(d_{\rm conf})$ and comparing the (relative) confidence needs $O(1)$. So the additional time complexity is $O(d_{\rm freq}\log (d_{\rm freq})+d_{\rm freq}d_{\rm conf})$, which is smaller than the original complexity $O(d_{\rm freq}^2\log (d_{\rm freq}))$. Thus these operations will not burden the computation as well.

\subsection{Parameter Selection}
If computation resource is limited, it's vital to control the computation complexity.
From the analysis above, time complexity is at least $O(np)$ and space complexity is at least $O(p)$.
If $O(d_{\rm freq}^2)$ is smaller than $O(p)$, both time and space complexities keep unchanged when $d_{\rm freq}$ increases.
On the other hand, if $O(d_{\rm freq}^2)$ is larger than $O(p)$, complexities will increase with $d_{\rm freq}$.
So setting $d_{\rm freq}$=$O(\sqrt{p})$ is a good choice that will not carry too much computation cost while generating features as many as possible.

What's left is how to choose $d_{\rm conf}$ when $d_{\rm freq}$ is determined.
Keep in mind that $d_{\rm conf}$ is the number of association rules mined by ARAF, but may not be the number of features generated by ARAF.
This is because an antecedent may imply different consequences, thus different rules can generate the same feature.
It's not pleasant since the we can not control exactly how many features will be generated.
Notice that an antecedent can at most correspond to $|C|$ rules,
so we can ease the problem by setting the ratio of parameters $d_{\rm freq}/d_{\rm conf}\ge |C|$.
Then it can be guaranteed that the selected frequent itemsets contain at least $d_{\rm conf}$ association rules corresponding to different features, though some of which may not be selected at last.

When computation resource is sufficient, it's not wise to enlarge the parameters unlimitedly. Because selecting the rules with small support may overfit the training set, and the rules with small confidence can actually be a noise. 
It is reasonable to compare ARAF with one-hot encoding, since their behaviors on main effects are similar.
The difference is that ARAF may only expand the main effects partially, while some of the interactions are added to the input.
Thus we can run ARAF with small $d_{\rm freq}$ and $d_{\rm conf}$ as stated above, then merely attach interactive features to the one-hot encoded data since the main effects have already been added.

\section{Other Modifications}\label{sec:modification}
\subsection{Non-exhaustive Version}
Though Apriori is appealing due to its simplicity, its efficiency is not satisfactory. 
This shortcoming is inherited by ARAF.
As analyzed in Section~\ref{sec:computational_complexity}, the running time is linear with the number of samples and the number of features.
Thus for data sets with really large $n$ and $p$, ARAF can be rather slow. 
One can substitute Apriori with more advanced data mining algorithms, which will not influence the resulting interactions if thresholds for support and confidence are the same.
Another choice to speed up ARAF is random sampling.
If we select $n'$ samples from the database with replacement, define a binary variable $\chi_{s,i}$ to indicate whether an itemset $s$ occurs in the i-th sample and use $\hat{p}_s=\frac{1}{n’}\sum_{i=1}^{n'}\chi_{s,i}$ to estimate the true frequency $p_s$.
Then $\chi_{s,1}, \chi_{s,2},...,\chi_{s,n'}$ are independent random variables bounded by the interval [0, 1], and is thus sub-Gaussian with parameter 1/2.
According to the Hoeffding bound, we have

\begin{equation}
    P(|\hat{p}_s-p_s|\ge \epsilon)\le 2e^{-2n'\epsilon^2}.
\end{equation}

For two itemsets with frequency $p_1$ and $p_2$, where $p_1-p_2=\epsilon>0$.
Then $\hat{p}_1-\hat{p}_2\le 0$ implies $|p_1-\hat{p}_1|\ge \epsilon/2$ or $|p_2-\hat{p}_2|\ge \epsilon/2$.
Therefore we have
\begin{equation}
    \label{eq:mis_prob}
    \begin{aligned}
        P(\hat{p}_1-\hat{p}_2\le 0) &\le P(|p_1-\hat{p}_1|\ge \epsilon/2) + P(|p_1-\hat{p}_1|\ge \epsilon/2)\\
        &\le 4e^{-n'\epsilon^2/2}.
    \end{aligned}
\end{equation}
According to (\ref{eq:mis_prob}), if the frequency of an itemset is 5\% larger than another, by sampling 5000 instances, 
the probability of mistakenly identifying the more frequent one is at most 0.8\%.
We are interested in distinguishing frequent itemsets from the others, not obtaining the precise frequency.
Therefore we can efficiently select the frequent itemsets by subsampling,
especially when the difference of supports between the frequent and infrequent itemsets is large.

If we treat the original training set as a subsample of the behind distribution, the analysis also holds.
So the larger database, the more reliable rules, but the heavier computational cost.
Also we can conclude that ARAF is more efficient for the data sets where the number of features is smaller and the gap between supports of frequent and infrequent itemsets is wider.

Confidence can be estimated based on the approximate frequency, or calculated by a single pass through the whole database.

\subsection{Extension to Continuous Features}
All the discussions above are on discrete features, and we can not apply ARAF to continuous features directly.
But it's not difficult to transform a continuous feature to a discrete one, by splitting the range of the continuous feature into different intervals. 
There are a number of works that involve discretizing continuous features based on the classification target\cite{fayyad93, catlett91, dougherty95}. 
Similar to the method called ``Recursive Minimal Entropy Partitioning'' in some literature, we sort the numerical feature, and find thresholds leading to the largest information gain(IG)\cite{quinlan}:

\begin{equation}
    \label{ig}
    \begin{aligned}
        {\rm IG}(D, f, T)&=Ent(D)-\sum_{i=1}^{k}\frac{|D_i|}{|D|}Ent(D_i)
    \end{aligned}
\end{equation}
where $D$ is the database, $f$ is a continuous feature, $T$ are the thresholds that divide $f$ into $k$ intervals, $D_i$ is a subset of D corresponding to the i-th interval of $f$.

We adopt the number of intervals after discretizing, namely $k$ instead of minimal information gain as the stopping criterion. The reason is analogous to why we prefer $d_{\rm freq}$ and $d_{\rm conf}$ that we can control the computation complexity directly. And setting $k$ manually can ensure the assumption in Section 4.1 that it is a constant independent of the sample size $n$. After obtaining T, we use these thresholds to separate the feature.
The procedure of discretizing a continuous feature is presented in Algorithm \ref{split}.

\begin{algorithm}
    \caption{Discretize: discretize a continuous feature}
    \label{split}
    \begin{algorithmic}[1]
        \REQUIRE $f$: records of a continuous feature;\\
        $k$: the number of categories after discretizing;\\
        $l$: the number of quantiles for one searching.
        \ENSURE the discrete version of the input feature;\\
        \STATE{Initialize $S=\{[a_1, b_1]\}$,\\ where $a_1=\min (f)$, $b_1=\max (f)$}
        \FOR{i = 2: $k$}
            \STATE{Calculate $l$-quantile statistics of each interval in S}
            \STATE{Calculate the IG of splitting at the quantiles}
            \STATE{Select the quantile $q_i$ that leads to the largest IG, suppose $q_i\in [a_i, b_i]\in S$}
            \STATE{$S\leftarrow S\setminus \{[a_i, b_i]\}\cup\{[a_i, q_i], [q_i, b_i]\}$}
        \ENDFOR
        \STATE{Split $f$ according to which interval an element falls in}
        \STATE{Return the discretized feature}
    \end{algorithmic}
\end{algorithm}

After getting discretized features, ARAF can be applied. We call this procedure Extended-ARAF or EARAF for short.
Intuitively, we prefer smaller $k$ because it is convenient for human beings to comprehend,
but the performance of $k$ depends on the data and the model.
The choice of $k$ may have significant effects on the association rules found by ARAF, and it is hard to predict which value of $k$ will lead to good results. Therefore we have to select $k$ by grid search, which is unpleasant. How to deal with continuous features in a more graceful way needs to be further studied.

\section{Experiments}\label{sec:experiment}
By using ARAF for feature selection, we can save time and memory since only the useful features are reserved.
Also better results may appear because irrelevant features are prevented while some meaningful interactions are added.
When ARAF is used for adding interactive features to the data, it is expected to achieve better performance since it can make use of interactions explicitly.

Here are some numerical simulations to illustrate these advantages. 
After obtaining the new features, We use logistic regression(LR) and multi-layer perceptron(MLP) as classifiers to test the effectiveness of ARAF in this section except Section~6.4. 
The MLP has two hidden layers, and each layer consists of 30 hidden units. To avoid overfitting, LASSO\cite{Tibshirani96} with penalty parameter 1 is used as a regularizer for LR, and MLP is regularized by early stopping\cite{bishop95, Sjoberg94}.

\subsection{Synthetic Data}
\begin{figure}[!t]
    \centering
    \includegraphics[width=2.5in]{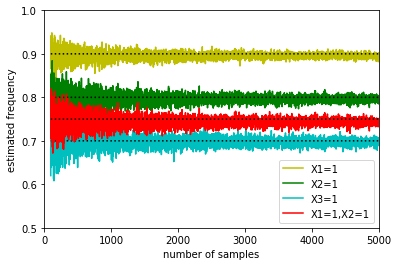}
    \caption{Estimated frequency of the frequent itemsets of the Synthetic Data through non-exhaustive ARAF.}
    \label{fig:frequency}
\end{figure}

\begin{figure}[!t]
    \centering
    \subfloat[Running time of different n']{\includegraphics[width=2.5in]{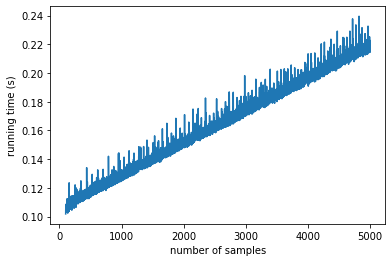}
    \label{fig:running_time_n}}
    \hfil
    \subfloat[Running time of different p]{\includegraphics[width=2.5in]{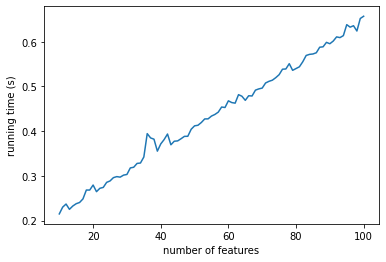}
    \label{fig:running_time_p}}
    \caption{Running time of different n' (a) and p (b) for non-exhaustive ARAF.}
    \label{fig:running_time}
\end{figure}

First we show that the non-exhaustive ARAF can find frequent itemsets efficiently.
A data set consisting 10000 instances and $p$ binary features are generated, where
$X_1$ has a Bernoulli(0.9) distribution;
$X_2$ is a variable dependent on $X_1$ such that $P(X_2\!=\!1|X_1\!=\!1)=75/90$, $P(X_2\!=\!1|X_1\!=\!0)=0.5$;
$X_3$ has a Bernoulli(0.7) distribution.
$X_4,...,X_p$ are i.i.d. Bernoulli(0.5) variables.
All the instances are labeled as 1 since we only care about whether the frequent itemsets can be found and whether the estimated frequency are accurate.
With $d_{\rm freq}$=5 and $p$=10, we tested the non-exhaustive ARAF with subsample size from 100 to 5000, and found that the 4 most frequent itemsets are always in the resulting sets. 
The estimated frequency for the frequent itemsets are shown in Fig. \ref{fig:frequency}, and the running time is exhibited in 
Fig. \ref{fig:running_time_n}.
We also test the running time of the non-exhaustive ARAF for different numbers of features. 
We set $n'=5000$ and $p$ from 10 to 100, and the running time is shown in Fig. \ref{fig:running_time_p}.
As expected, the estimated frequency converge to its true value, 
and the running time is almost linear with $n'$ and $p$.

Then we design two different data sets to illustrate the advantages and limitations of ARAF families according to their characteristics.

For the first case, 99 features, \{$X_1$, $X_2$, ..., $X_{99}$\}, are generated. $X_1$ is a Bernoulli(0.3) variable and $X_2$, $X_3$, ..., $X_{99}$ are i.i.d. Bernoulli(0.5) random variables. Label $Y$ is assigned 0, 1 or 2 according to (\ref{label_syn}).

\begin{equation}
    \label{label_syn}
    Y = \left\{
    \begin{aligned}
    0 & , & {X_1=0 } \\
    1 & , & {X_1=1, X_2X_3=0} \\
    2 & , & {X_1=1, X_2X_3=1}
    \end{aligned}
    \right.
\end{equation}

Thus only $X_1$, $X_2$ and $X_3$ are responsible for the label, while all the other features are irrelevant. And these settings lead to a highly unbalanced data set, where 70 percents of the instances are labeled as 0, 22.5 percents are labeled as 1 and only 7.5 percents are in the last class. Then we add some noise to the label, by randomly selecting 5 percents of the samples and setting their labels as 0, 1 or 2 with equal probability. We call this data set ``S1''. 

A harder case is when there are some redundant features. All other settings keep the same as S1, except that $X_{98}$ and $X_{99}$ always have value 1 in this case. We call this data set ``S2''.

Setting $d_{\rm freq}$=45 and $d_{\rm conf}$=5 for both S1 and S2, we generate the data sets consisting of 1000 instances and run the algorithms.
The experiments are repeated 100 times. The 5 most frequent association rules found by Algorithms \ref{araf_fs}, \ref{araf_ub} and \ref{araf_reluctant}(denoted by ALG4, ALG5 and ALG6, respectively) are shown in Table \ref{rule_syn}, with their number of occurrence following the comma.
We use ``($X_i$=$x_i$)$\to$c'' to stand for the rule ``If $X_i$=$x_i$, then Y=c'', and ``($X_i$=$x_i$, $X_j$=$x_j$)$\to$c'' for ``If $X_i$=$x_i$ and $X_j$=$x_j$, then Y=c''.
As expected, Algorithm \ref{araf_fs} tends to find rules about major class, which usually have larger supports.
And Algorithm \ref{araf_ub} can not bring out a satisfactory result for S2 because some useful rules are crowded out by redundant ones. 
And we can conclude that even when the number of the generated rules is limited, Algorithm \ref{araf_reluctant} can find rules for different classes as many as possible.

\begin{table*}[!t]
    \renewcommand{\arraystretch}{1.3}
    \caption{Rules for Synthetic Data}
    \label{rule_syn}
    \centering
    \begin{tabular}{c||c|ccccc}
        \hline
         & method & Rule1 & Rule2 & Rule3 & Rule4 & Rule5 \\
        \hline
        \multirow{3}{*}{S1} & ALG4 & ($X_{1}$=0,$X_{82}$=1)$\to$0, 8 & ($X_{1}$=0,$X_{78}$=0)$\to$0, 8 & ($X_{1}$=0,$X_{65}$=0)$\to$0, 7 & ($X_{1}$=0,$X_{27}$=0)$\to$0, 7 & ($X_{1}$=0,$X_{35}$=1)$\to$0, 7 \\
        & ALG5 & ($X_{1}$=1,$X_{3}$=0)$\to$1, 97 & ($X_{1}$=1,$X_{2}$=0)$\to$1, 96 & ($X_{1}$=0)$\to$0, 18 & ($X_{1}$=1,$X_{2}$=1)$\to$2, 7 & ($X_{1}$=1,$X_{3}$=1)$\to$2, 6 \\
        & ALG6 & ($X_{1}$=1,$X_{3}$=0)$\to$1, 97 & ($X_{1}$=0)$\to$0, 97 & ($X_{1}$=1,$X_{2}$=0)$\to$1, 96 & ($X_{1}$=1,$X_{2}$=1)$\to$2, 88 & ($X_{1}$=1,$X_{3}$=1)$\to$2, 81 \\
        \hline
        \multirow{3}{*}{S2} & ALG4 & ($X_{1}$=0)$\to$0, 32 & ($X_{1}$=0,$X_{99}$=1)$\to$0, 26 & ($X_{1}$=0,$X_{98}$=1)$\to$0, 21 & ($X_{1}$=0,$X_{69}$=1)$\to$0, 6 & ($X_{1}$=0,$X_{22}$=0)$\to$0, 6\\
        & ALG5 & ($X_{1}$=1,$X_{2}$=0)$\to$1, 86 & ($X_{1}$=1,$X_{3}$=0)$\to$1, 74 & ($X_{1}$=0)$\to$0, 74 & ($X_{1}$=0,$X_{98}$=1)$\to$0, 71 & ($X_{1}$=0,$X_{99}$=1)$\to$0, 67 \\
        & ALG6 & ($X_{1}$=0)$\to$0, 100 & ($X_{1}$=1,$X_{3}$=1)$\to$2, 88 & ($X_{1}$=1,$X_{2}$=0)$\to$1, 86 & ($X_{1}$=1,$X_{2}$=1)$\to$2, 84 & ($X_{1}$=1,$X_{3}$=0)$\to$1, 74 \\
        \hline
    \end{tabular}
\end{table*}

We also use the original features and the features generated by the algorithms to train an LR and an MLP for each above-mentioned data set. The average results of 100 trials with standard error in the parentheses are listed in Table \ref{table_syn}, where ``ACC'' stands for accuracy and ``Logloss'' represents cross entropy.
We can make a summary that the algorithms, especially Algorithm \ref{araf_reluctant}, can find useful features, so memory and time can be saved while the performance is improved.
ARAF may not mine the rules from which the data sets are generated, since the intrinsic rules can not be expressed by 1-item or 2-item antecedents. But it can find out almost all the main effects and interactions responsible for the target, even with a small number of association rules. 

\begin{table}[!t]
    \renewcommand{\arraystretch}{1.3}
    \caption{Results of Synthetic Data}
    \label{table_syn}
    \centering
    \begin{tabular}{c|c|cc|cc} 
        \hline
        \multirow{2}{*}{} & \multirow{2}{*}{method} & \multicolumn{2}{c}{LR} & \multicolumn{2}{c}{MLP}\\
        \cline{3-6}
        & & logloss & ACC(\%) & logloss & ACC(\%) \\
        \hline
        \multirow{4}{*}{S1} & origin & 0.274(0.076) & 95.1(1.8) & 0.323(0.062) & 90.7(2.1) \\
        & ALG4 & 0.407(0.063) & 87.5(2.5) & 0.399(0.060) & 87.5(2.2) \\
        & ALG5 & 0.232(0.075) & 94.0(3.9) & 0.223(0.070) & 93.1(4.3) \\
        & ALG6 & 0.180(0.054) & 96.7(1.4) & 0.176(0.053) & 96.5(1.4) \\
        \hline
        \multirow{4}{*}{S2} & origin & 0.278(0.076) & 95.0(1.8) & 0.311(0.073) & 90.8(2.3) \\
        & ALG4 & 0.374(0.076) & 88.2(2.3) & 0.362(0.071) & 88.7(2.2) \\
        & ALG5 & 0.215(0.070) & 93.4(4.1) & 0.211(0.069) & 93.8(3.8) \\
        & ALG6 & 0.183(0.050) & 96.4(1.4) & 0.173(0.052) & 96.8(1.3) \\
        \hline
    \end{tabular}
\end{table}

\subsection{Industrial Data}
\begin{figure}[!t]
    \centering
    \includegraphics[width=2.5in]{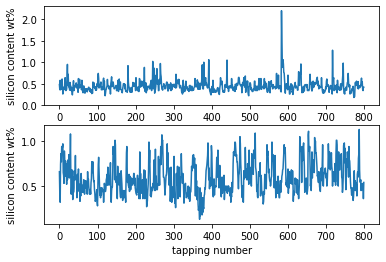}
    \caption{Evolution of the hot metal silicon content in BF-a and BF-b.}
    \label{si}
\end{figure}

The data in this section is collected from two blast furnaces, denoted by BF-a and BF-b. 
For BF-a, silicon content lower than 0.3736 will be regarded as low, higher than 0.8059 will be seen as high, otherwise is proper. Similarly, the corresponding thresholds for BF-b are 0.4132 and 0.8251\cite{Luo11}.
The target is to predict whether the silicon content of hot metal is low, high or proper.
Some related variables as well as the silicon content of the past are provided.
In total, there are 27 features for BF-a and 86 for BF-b, all of which are continuous.
And there are 800 instances in each data set. The evolution of the hot metal silicon content in BF-a and BF-b is illustrated in Fig. \ref{si}, from which we can see that the behavior of silicon content is very complicated.

As the separation in \cite{gao14}, we use the first 600 samples for training, the next 100 samples for validation and the last 100 samples for test. The distribution of the labels in BF-a and BF-b are presented in Table \ref{label_bf}. We can see that the data is awfully unbalanced, and the distribution is shifting over time.

\begin{table}[!t]
    \renewcommand{\arraystretch}{1.3}
    \caption{Distribution of instances in BF-a and BF-b}
    \label{label_bf}
    \centering
    \begin{tabular}{c|c|ccc} 
        \hline
        \multirow{2}{*}{Blast Furnace} & \multirow{2}{*}{data set} & \multicolumn{3}{c}{Silicon content}\\
        \cline{3-5}
        & & low & proper & high\\
        \hline
        \multirow{3}{*}{BF-a} & Training & 155 & 428 & 17 \\
        & Validation & 31 & 68 & 1 \\
        & Test & 30 & 68 & 2 \\
        \hline
        \multirow{3}{*}{BF-b} & Training & 104 & 427 & 69 \\
        & Validation & 22 & 73 & 5 \\
        & Test & 19 & 74 & 7 \\
        \hline
    \end{tabular}
\end{table}

We show the procedure of training LR on data set BF-a with logloss as the criterion in detail for illustration, and simply give the results of other situations.

First we have to determine the parameters, including $k$, $d_{\rm freq}$ and $d_{\rm conf}$.
It's computationally expensive to optimize three parameters at the same time by grid search.
Frequent itemsets keep unchanged if $d_{\rm freq}$ is fixed, so increasing $d_{\rm conf}$ need only select different number of rules that are most confident from a min-heap. 
With the consideration of this fact, we fix $d_{\rm freq}$ beforehand. For different $k$, we first split the continuous features into $k$ parts and mine the $d_{\rm freq}$ most frequent itemsets, then calculate the relative confidence of each rule and by which sort the rules. Then for different $d_{\rm conf}$, we simply select the most relatively confident $d_{\rm conf}$ rules without additional operations. 

\begin{figure}[!t]
    \centering
    \includegraphics[width=2.5in]{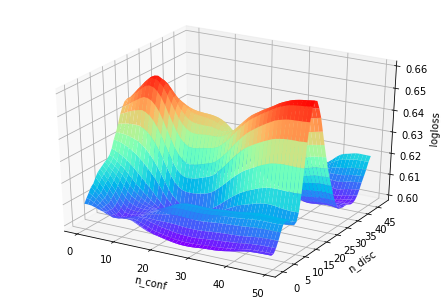}
    \caption{Logloss of LR on BF-a.}
    \label{logloss_lr_lg}
\end{figure}

Setting $d_{\rm freq}$=150, the range of $k$ is [2, 49], the range of $d_{\rm conf}$ is [0, 49]. 
The results on the validation set is very noisy, and we can not have a moderate overview of how the logloss changes with different parameters without further processing. Luckily there are numerous methods for smoothing, such as box filter, median filter\cite{Huang79}, Gaussian filter or bilateral filter\cite{Tomasi98}. We adopt Gaussian filter to smooth the results and 
the performance of LR with different $k$ and $d_{\rm conf}$ on the validation set is shown in Fig. \ref{logloss_lr_lg}.

We can see that logloss is generally smaller when $k$ is sufficiently small or sufficiently large, and LR performs best when ($k, d_{\rm conf}$) is near (4, 25).

With $k$=4, $d_{\rm freq}$=150 and $d_{\rm conf}$=25, we train LR on the union of training set and validation set, and calculate the logloss on the test set. The results are listed in Table \ref{table_result}, from which we can see that EARAF makes some improvements.

The procedure of dealing with other combinations of data sets, models or criteria is similar. 
The parameters for each situation are presented in Table \ref{table_parameters}. We can conclude that MLP can usually take care of more features generated from rules than LR. Adding features is likely to be beneficial, at least not harmful to MLP. However, LR usually performs best when $d_{\rm conf}$ is appropriately small. In the case where LR is trained with logloss as criteria on BF-b, it surprisingly rejects all the rules. This may be attributed to distribution shifting, thus adding rules makes it easier for LR to overfit the training set. But even though without rules, discretizing continuous feature comes into its own and makes some improvements, which coincides with the conclusion in \cite{dougherty95}. Another interesting fact is that the same model achieves its optimal performance with different parameters when using different criteria. This may due to the unbalancedness of the data. Force the model to pay more attention to minor class may reduce the confidence of classifying an sample as major class. The confusion matrices of LR on BF-a with ($k$=0, $d_{\rm conf}$=0)(original features), ($k$=4, $d_{\rm conf}$=25)(the best parameters for logloss) and ($k$=2, $d_{\rm conf}$=10)(the best parameters for accuracy) may illustrate this conjecture, which are shown in Table \ref{confusion_matrix}. We can see that LR with (0, 0) classifies most of the samples as ``proper'', while LR with (4, 25) correctly recognizes more samples of ``low'' at the expense of misclassifying some ``proper'' instances. LR with (2, 10) reaches a slightly higher accuracy, but also has a larger logloss(0.6328 indeed) than LR with (4, 25).

The results of each situation are exhibited in Table \ref{table_result}.
By comparing the results of EARAF and original features, we can summarize that EARAF make some improvements.
And EARAF can also obtain better results than those reported in \cite{gao14}, where accuracy is 68\%  on BF-a and 79\% on BF-b.

\begin{table}[!t]
    \caption{Parameters for Different Situations}
    \label{table_parameters}
    \centering
    \begin{tabular}{c|c||cc|cc} 
        \hline
        \multirow{2}{*}{data set} & \multirow{2}{*}{criterion} & \multicolumn{2}{c|}{LR} & \multicolumn{2}{c}{MLP}\\
        \cline{3-6}
         & & $k$ & $d_{\rm conf}$ & $k$ & $d_{\rm conf}$\\
        \hline
        \multirow{2}{*}{BF-a} & logloss & 4 & 25 & 3 & 40\\
                           & ACC & 2 & 10 & 15 & 40\\
        \hline
        \multirow{2}{*}{BF-b} & logloss & 30 & 0 & 8 & 20\\
                           & ACC & 30 & 20 & 3 & 40\\
        \hline
    \end{tabular}
\end{table}

\begin{table}[!t]
    \caption{Confusion matrixes of LR on BF-a}
    \label{confusion_matrix}
    \centering
    \begin{tabular}{c||c|ccc} 
        \hline
        ($k$, $d_{\rm conf}$) & label & low & proper & high \\
        \hline
        \multirow{3}{*}{(0, 0)} & low & 5 & 25 & 0 \\
        & proper & 0 & 68 & 0 \\
        & high & 0 & 2 & 0 \\
        \hline
        \multirow{3}{*}{(4, 25)} & low & 9 & 21 & 0 \\
        & proper & 3 & 64 & 1 \\
        & high & 1 & 1 & 0 \\
        \hline
        \multirow{3}{*}{(2, 10)} & low & 10 & 20 & 0 \\
        & proper & 3 & 64 & 1 \\
        & high & 1 & 1 & 0 \\
        \hline
    \end{tabular}
\end{table}

\begin{table}[!t]
    \caption{Results of Different Situations}
    \label{table_result}
    \centering
    \begin{tabular}{c|c||cc||cc} 
        \hline
        \multirow{2}{*}{data set} & \multirow{2}{*}{method} & \multicolumn{2}{c||}{LR} & \multicolumn{2}{c}{MLP}\\
        \cline{3-6}
         & & logloss & ACC(\%) & logloss & ACC(\%)\\
        \hline
        \multirow{2}{*}{BF-a} & origin & 0.6419 & 73 & 0.6217 & 71\\
                           & EARAF & 0.5817& 74 & 0.5241 & 74\\
        \hline
        \multirow{2}{*}{BF-b} & origin & 0.4916 & 77 & 0.6783 & 77\\
                           & EARAF & 0.4484 & 83 & 0.5776 & 79\\
        \hline
    \end{tabular}
\end{table}

\subsection{Public data sets}
We also conduct experiments on classical public data sets.
We choose four data sets from the UCI Machine Learning repository\cite{uci}.

The first data set, denoted by ``Adult'', is extracted by Barry Becker from the 1994 Census database. It contains 48842 instances, each has 6 continuous and 8 discrete features, and the prediction task is to determine whether a person makes over 50K a year. 

The second data set named ``Heart Disease'' and ``HD'' for short, collected by Robert Detrano et al., consists of 303 examples and 13 features, 9 of which are discrete. The target is to distinguish presence of heart disease in the patient (values 1,2,3,4) from absence (value 0).

The third data set, named ``Default of Credit Card Clients''\cite{Yeh09} and ``DCCC'' for short. There are 30000 examples and 23 features in the data. 10 of the features are discrete while the other 13 are continuous. We need to predict the default payment (Yes = 1, No = 0).

The last data set is ``Car Evaluation'', ``CE'' for short, which contains 1728 samples and 6 categorical features are giving to classify the instance as ``unacc'', ``acc'', ``good'', or ``vgood''. Thus it's a multiclass classification task \cite{Bohanec88}.

\begin{table*}[!t]
    \caption{Results of UCI data sets}
    \label{table_uci}
    \centering
    \begin{tabular}{c||c|ccc|ccc} 
        \hline
        \multirow{2}{*}{data set} & \multirow{2}{*}{method} & \multicolumn{3}{c|}{LR} & \multicolumn{3}{c}{MLP}\\
        \cline{3-8}
        & & logloss & ACC(\%) & AUC & logloss & ACC(\%) & AUC\\
        \hline
        \multirow{4}{*}{Adult} & LABEL & 0.3776(0.0150) & 82.85(0.68) & 0.7537(0.0079) & 0.3182(0.0069) & 85.13(0.27) & 0.7599(0.0047)\\
        & EARAF-L & \textbf{0.3228(0.0077)} & \textbf{85.04(0.42)} & \textbf{0.7621(0.0034)} & \textbf{0.3161(0.0091)} & \textbf{85.20(0.28)} & \textbf{0.7635(0.0126)}\\
        \cline{2-8}
        & ONEHOT & 0.3184(0.0066) & 85.26(0.33) & 0.7727(0.0078) & 0.3091(0.0063) & 85.73(0.32) & 0.7718(0.0075)\\
        & EARAF-O & \textbf{0.3124(0.0081)} & \textbf{85.65(0.43)} & \textbf{0.7743(0.0061)} & \textbf{0.3071(0.0071)} & \textbf{85.86(0.32)} & \textbf{0.7736(0.0117)} \\
        \hline
        \multirow{4}{*}{HD} & LABEL & 0.4167(0.0327) & 81.82(5.01) & 0.8136(0.0517) & 0.4849(0.0388) & 79.87(3.79) & 0.7736(0.0277)\\
        & EARAF-L & \textbf{0.3861(0.0400)} & \textbf{83.15(4.53)} & \textbf{0.8295(0.0481)} & \textbf{0.3814(0.0420)} & \textbf{84.16(2.02)} & \textbf{0.8236(0.0456)}\\
        \cline{2-8}
        & ONEHOT & 0.3836(0.0279) & 84.80(4.01) & 0.8442(0.0437) & 0.3778(0.0412) & 85.46(3.41) & 0.8437(0.0212)\\
        & EARAF-O & 0.3946(0.0352) & 83.48(4.61) & 0.8320(0.0488) & \textbf{0.3678(0.0430)} & \textbf{85.47(3.43)} & \textbf{0.8563(0.0320)}\\
        \hline
        \multirow{4}{*}{DCCC} & LABEL & 0.4712(0.0083) & 79.98(0.46) & 0.5843(0.0171) & 0.4427(0.0067) & 81.64(0.48) & 0.6255(0.0162)\\
        & EARAF-L & \textbf{0.4435(0.0045)} & \textbf{81.29(0.24)} & \textbf{0.6376(0.0058)} & \textbf{0.4401(0.0033)} & 81.27(0.46) & \textbf{0.6420(0.0096)}\\
        \cline{2-8}
        & ONEHOT & 0.4368(0.0041) & 81.96(0.29) & 0.6527(0.0036) & 0.4315(0.0041) & 82.04(0.33) & 0.6610(0.0041)\\
        & EARAF-O & \textbf{0.4352(0.0039)} & \textbf{82.04(0.30)} & \textbf{0.6533(0.0038)} & 0.4318(0.0039) & 81.97(0.33) & 0.6580(0.0037)\\
        \hline
        \multirow{4}{*}{CE} & LABEL & 0.4726(0.0376) & 79.34(2.03) & - & 0.1443(0.0173) & 94.97(0.86) & - \\
        & EARAF-L & \textbf{0.2285(0.0200)} & \textbf{91.20(1.35)} & - & \textbf{0.1146(0.0341)} & \textbf{96.01(0.69)} & - \\
        \cline{2-8}
        & ONEHOT & 0.2431(0.0079) & 88.83(1.08) & - & 0.0306(0.0273) & 98.50(0.99) & - \\
        & EARAF-O & \textbf{0.2071(0.0174)} & \textbf{91.72(1.40)} & - & 0.0668(0.0169) & \textbf{98.67(0.29)} & - \\
        \hline
    \end{tabular}
\end{table*}

All the data sets contain both continuous features and discrete features except CE. Therefore we first discrete the continuous features and add the discretized features to the original data, as stated in Section~\ref{sec:modification}. The parameter $k$ is decided by grid search from 2 to 10, scored by the criterion on a validation set.

To spare time and show the robustness of EARAF, we fix $d_{\rm freq}$ and $d_{\rm conf}$ in advance this time.
As suggested in Section 4, we set $d_{\rm freq}$=$5|C|\lfloor\sqrt{p}\rfloor$, $d_{\rm conf}$=$5\lfloor\sqrt{p}\rfloor$. And we adopt two different ways to make use of the generated features. The first one, which named ``EARAF-L'', is adding all the generated features to the label-encoded data. This setting is a simulation of the case when computation resource is limited that one-hot encoding is infeasible. The next is that first one-hot encode the discrete features and then only attach the interactive features, which we call ``EARAF-O''. This corresponds to adding interactive features. The effects of EARAF-L, EARAF-O as well as label encoding and one-hot encoding are tested. 

As earlier experiments, LR and MLP are used for prediction while logloss and accuracy are used for model evaluation.
Noticing that the label of most of the data sets is binary, so we further adopt AUC\cite{Spackman89} as a new criterion.
To be more convincing, 5-fold cross validation is applied for every case. The results of each data set are shown in Table \ref{table_uci}, in which the results are shown in bold if EARAF-L outperforms label encoding or EARAF-O outperforms one-hot encoding.

As can be seen from the table, EARAF-L almost always has better performance than label-encoding and makes remarkable improvements. It is encouraging since only a small number of features are added. And EARAF-O usually outperforms one-hot encoding, which means adding interactive features makes sense.
However, there are some exceptions. For example, LR on data set HD gets worse results if EARAF-O is applied, which may be caused by overfitting.

Another observation is LR usually benefits more from EARAF then MLP, and a possible explanation is that MLP has already found useful interactions itself. Though EARAF sometimes does little to help MLP, such as the case training MLP on data set DCCC, taking the interpretability and simplicity of LR into account, EARAF is still of great realistic significance.

\subsection{Comparison with Existing Methods}
We also apply ARAF on the two data sets used in \cite{shah16}.
The first is Communities and Crime Unnormalized Data Set, ``CCU'' for short, 
which contains crime statistics for the year 1995 obtained from FBI data, and national census data from 1990.
We take violent crimes per capita as our response, which makes it a regression task.
We preprocess the data by the procedure in \cite{shah16}.
This leads to a data set consisting of 1903 observations and 101 features.

The second data set is ``ISOLET'', which consists of 617 features based on the speech waveforms generated from utterances of each letter of the English alphabet.
We consider classification on the notoriously challenging E-set consisting of the letters ``B'', ``C'', ``D'', ``E'', ``G'', ``P'', ``T'', ``V'' and ``Z''.
And finally we have 2700 observations spread equally among 9 classes.

We use a Lasso as the base regression procedure, 
and penalised multinomial regression for the classification example.
The regularization coefficient is determined by 5-fold cross-validation.
To evaluate the procedures, we randomly select 2/3 for training and the remaining 1/3 for testing. 
This is repeated 200 times for each data set. 
The criterion for regression model is mean square error, and misclassification rate is used for classification.
All the settings are exactly the same as in \cite{shah16}, 
except we use $l_2$-regularizer to penalise the regression model instead of group Lasso.
This is because we don't know how the authors grouped the features in \cite{shah16} and it's time-consuming to apply group Lasso.

To apply ARAF, the numerical response of ``CCU'' is split into 5 categories by quantiles to obtain a discrete version, and the continuous features are then discretized by Algorithm \ref{split} with $k$=5. 
Setting $d_{\rm freq}$=2500 and $d_{\rm conf}$=1225, we add $X_iX_j$ as an interactive feature to the input if there is a rule with antecedent ($X_i$=$x_i$, $X_j$=$x_j$) for some $x_i$ and $x_j$ in the resulting rule sets.
The results of our models with and without ARAF are shown in Table \ref{tab:ccu_isolet}, labeled as ``ARAF*'' and ``Main*''.
We also listed the results reported in \cite{shah16}, including base procedures (``Main''), iterated Lasso fits (``Iterated''), Lasso following marginal screening for interactions (``Screening''), Random Forests \cite{Breiman01}, hierNet \cite{bien13} and MARS \cite{Friedman91}.
For data set ``CCU'', our base model outperforms the one in \cite{shah16}, which may caused by a better penalty parameter.
And the Lasso with ARAF leads to comparable or better result when compared to existing algorithms. 
As for ``ISOLET'', the result of our base model is not as good as the one in \cite{shah16}, this is not surprising since we simply use $l_2$-regularizer while Shah et al. adopt group Lasso to penalise the model.
But we can see that ARAF can run on this data set and lead to a good improvement, while some existing methods such as Screening, hierNet, MARS are inapplicable.
We think this could be an evidence of ARAF's efficiency.

\begin{table}[!t]
    \caption{Results of CCU and ISOLET}
    \label{tab:ccu_isolet}
    \centering
    \begin{tabular}{c||c|c} 
        \hline
        \multirow{2}{*}{method} & \multicolumn{2}{c}{ERROR}\\
        \cline{2-3}
        &  Communities and crime & ISOLET\\
        \hline
        Main & $0.414(6.5\times 10^{-3}$) & $0.0641(4.7\times 10^{-4}$)\\
        Iterate & $0.384(5.9\times 10^{-3}$) & $0.0641(4.7\times 10^{-4}$)\\
        Screening & $0.390(7.8\times 10^{-3}$) & - \\
        Backtracking & $0.365(3.7\times 10^{-3}$) & $0.0563(4.5\times 10^{-4}$)\\
        Random Forest & $0.356(2.4\times 10^{-3}$) & $0.0837(6.0\times 10^{-4}$)\\
        hierNet & $0.373(4.7\times 10^{-3}$) & - \\
        MARS & $5580.586(3.1\times 10^{3}$) & - \\
        Main* & $0.404(5.7\times 10^{-3})$ & $0.0730(5.5\times 10^{-4}$) \\
        ARAF* & $0.373(5.3\times 10^{-3}$) & $0.0677(5.0\times 10^{-4}$) \\
        \hline
    \end{tabular}
\end{table}

\section{Conclusion}\label{sec:conclusion}
Inspired by association rule mining, we propose a method that mines useful main or interaction effects from data, 
Instead of simply employing Apriori to mine association rules, we modify the algorithm for several practical concerns, such as (1) using the number of frequent sets and confident rules as parameters, which makes it more robust and easier to control the computation complexity;
(2) mining frequent sets for different target class separately and selecting rules with largest relative confidence instead of confidence to obtain rules for minor class;
(3) giving priority to main effects, thus redundant rules can be prevented;
(4) speeding up the algorithm by random sampling, thus it can be used for large data sets;
(5) extending the algorithm for continuous features by discretizing them. 
Also, we adopt a special data structure, namely min-heap, to optimize the computation complexity. We analyze the time and space complexity to show the efficiency of our algorithm, based on which we give some advice on parameter selection. Finally we conduct a number of experiments on synthetic, industrial and public data sets. 
Regardless of whether the data set is large or small, the original features are discrete or continuous, the classification task is binary or multiclass, the model is shallow or deep, 
the results show that EARAF can lead to improvement to some extent in most cases.\\

Future work mainly concentrate on the following directions.
The first is to adopt a more efficient association rule mining algorithm.
And if the running time can be shortened, higher-order interactions can be mined in the framework of ARAF.
The next is to find a more efficient approach to tune the parameters.
Though we have provided some idea about parameter selection based on computation complexity, but can not totally get rid of grid search, which is very time-consuming.
Also we hope to find a more natural way to deal with the continuous features.
Another target is to provide the proposed algorithm some theoretical supports. 
For example, we want to give a solid theoretical foundation for the information principle under some assumptions, or show that adding interactions is beneficial theoretically.

\ifCLASSOPTIONcompsoc
  \section*{Acknowledgments}
  This work was supported by the National Nature Science Foundation of China under Grant No. 12071428 and 11671418, and the Zhejiang Provincial Natural Science Foundation of China under Grant No. LZ20A010002.
\else
  \section*{Acknowledgment}
\fi

\bibliographystyle{IEEEtran}
\bibliography{IEEEabrv, ref}

\begin{IEEEbiography}[{\includegraphics[width=1in,height=1.25in,clip,keepaspectratio]{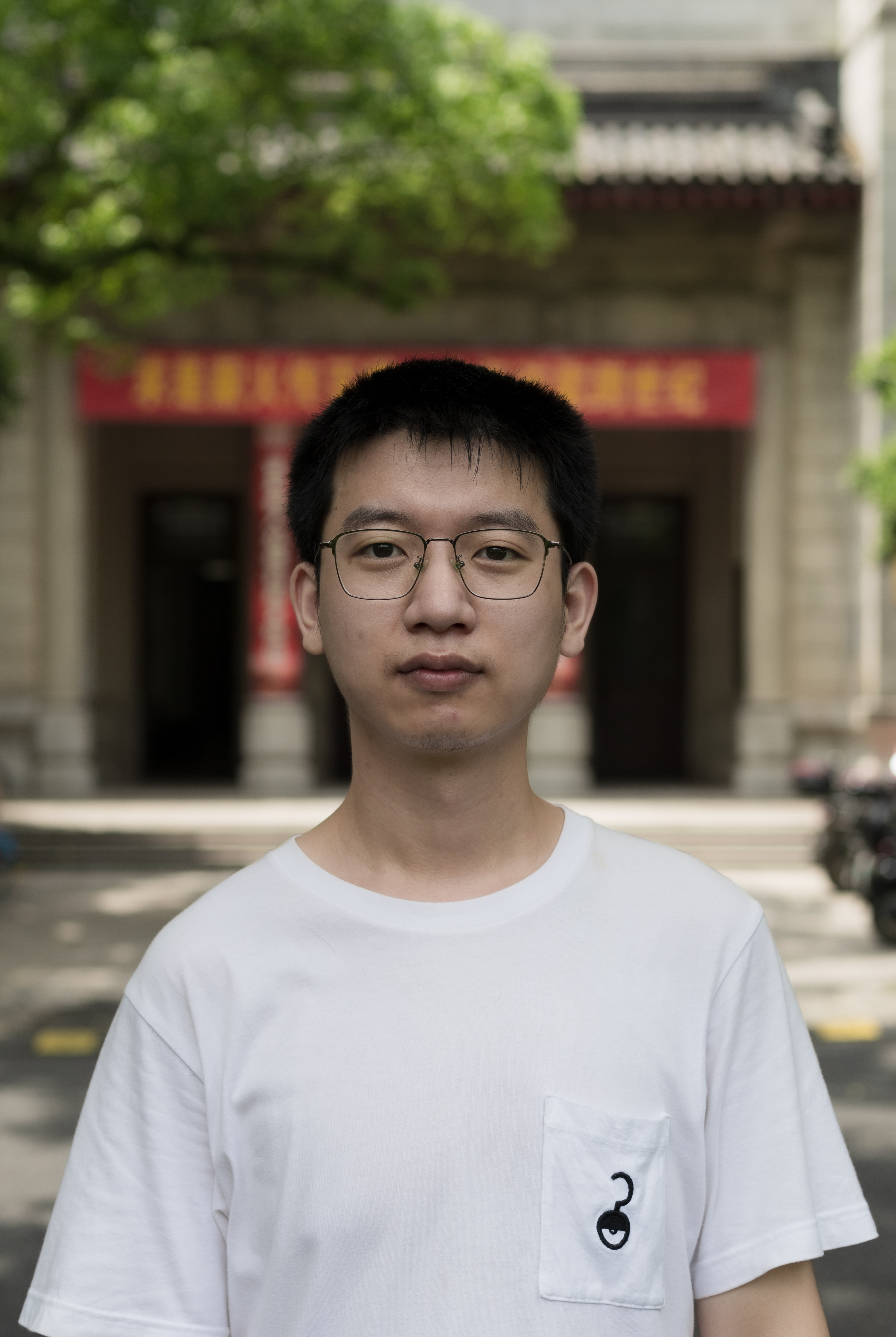}}]{Qiuqiang Lin}
received the B.S. degrees in Mathematics and Applied Mathematics from Zhejiang University, China, in 2017. He is currently working towards the Ph.D. degree in operational research and cybernetics at Zhejiang University.
His research interests are in the areas of machine learning applications and machine learning theory.
\end{IEEEbiography}

\begin{IEEEbiography}[{\includegraphics[width=1in,height=1.25in,clip,keepaspectratio]{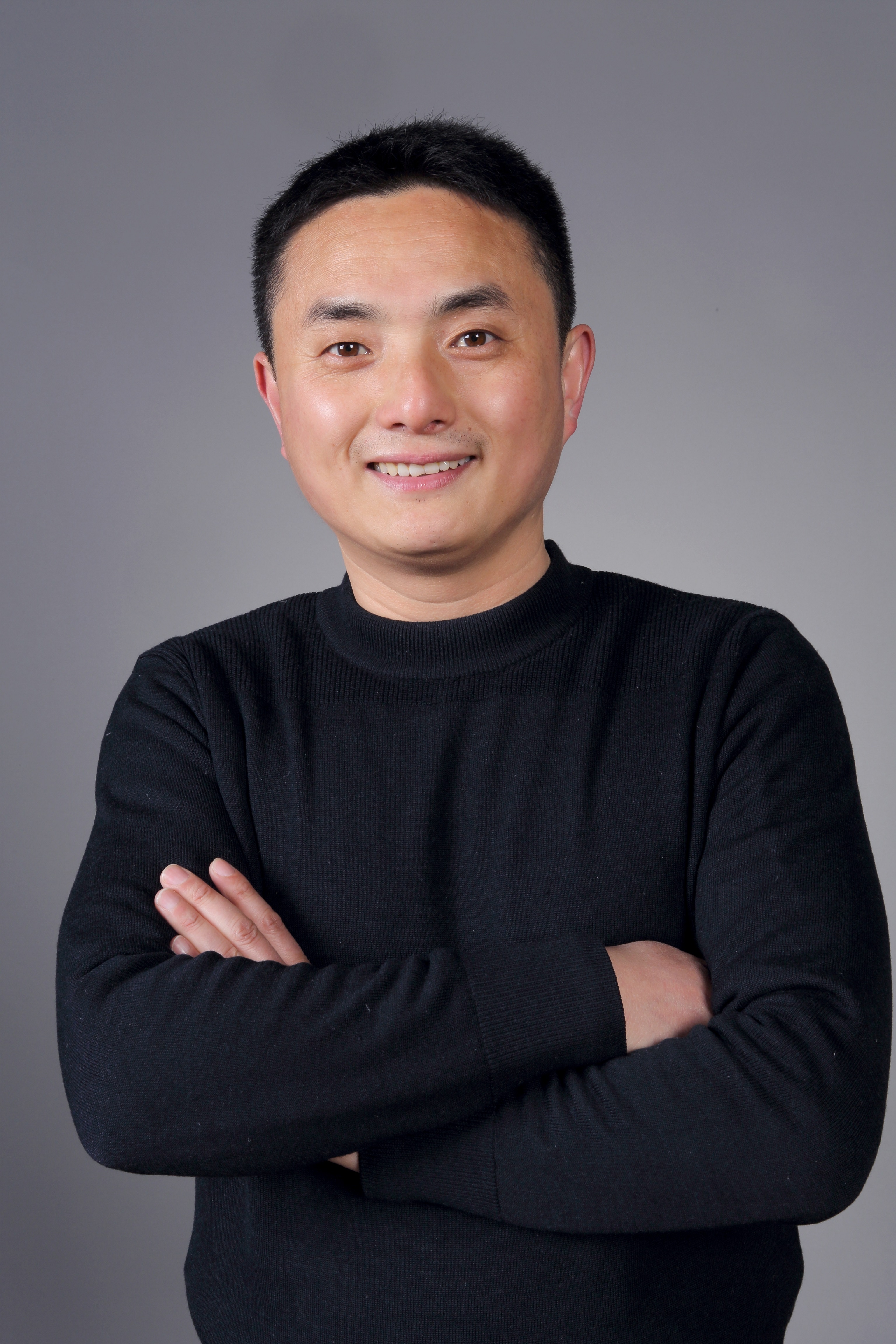}}]{Chuanhou Gao}
(M’09 SM’12) received the B.Sc. degrees in Chemical Engineering from Zhejiang University of Technology, China, in 1998, and the Ph.D. degrees in Operational Research and Cybernetics from Zhejiang University, China, in 2004. From June 2004 until May 2006, he was a Postdoctor in the Department of Control Science and Engineering at Zhejiang University.\\
Since June 2006, he has joined the Department of Mathematics at Zhejiang University, where he is currently a Professor. He was a visiting scholar at Carnegie Mellon University from Oct. 2011 to Oct. 2012. His research interests are in the areas of data-driven modeling, control and optimization, chemical reaction network theory and thermodynamic process control. He was a guest editor of IEEE Transactions on Industrial Informatics, ISIJ International and Journal of Applied Mathematics, and an editor of
Metallurgical Industry Automation from May 2013.
\end{IEEEbiography}

\end{document}